\documentclass[a4paper, superscriptaddress, twocolumn, prl
]{revtex4-1}

\usepackage[colorlinks=true,citecolor=blue,linkcolor=blue,pdfhighlight =/O]{hyperref}
\usepackage{graphicx}
\usepackage{amsmath}
\usepackage{mathtools}
\usepackage{color}
\usepackage{soul}

\hypersetup{urlcolor=blue}

\begin{document}

% Use the \preprint command to place your local institutional report number in the upper righthand corner of the title page in preprint mode.
% Use the 'preprintnumbers' class option to override journal defaults  to display numbers if necessary
%\preprint{}

%\title{Electron Over-Heating Masks Shapiro Steps in Proximity Josephson Junctions}
%\title{Interplay of Electron Over-Heating and Shapiro Steps in a Josephson Junction}
\title{Interplay Between Electron Over-Heating and ac Josephson Effect}
% repeat the \author .. \affiliation  etc. as needed  \email, \thanks, \homepage, \altaffiliation all apply to the current author. Explanatory text should go in the []'s, actual e-mail address or url should go in the {}'s for \email and \homepage.
% Please use the appropriate macro foreach each type of information \affiliation command applies to all authors since the last
% \affiliation command. The \affiliation command should follow the other information
\author{A. De Cecco$^{1,2}$, K. Le Calvez$^{1,2}$, B. Sac\'ep\'e$^{1,2}$, C. B. Winkelmann$^{1,2}$ and H. Courtois$^{1,2}$}
%\thanks{}
\affiliation{$^{1}$Universit\'e Grenoble Alpes, Institut N\' eel, 25 Avenue des Martyrs, 38042 Grenoble, France}
\affiliation{$^{2}$CNRS, Institut N\' eel, 25 Avenue des Martyrs, 38042 Grenoble, France}

%Collaboration name if desired (requires use of superscriptaddress option in \documentclass). \noaffiliation is required (may also be used with the \author command).
%\collaboration can be followed by \email, \homepage, \thanks as well. \collaboration{}

\date{\today}

\begin{abstract}
We study the response of high-critical current proximity Josephson junctions to a microwave excitation. Electron over-heating in such devices is known to create hysteretic dc voltage-current characteristics. Here we demonstrate that it also strongly influences the ac response. The interplay of electron over-heating and ac Josephson dynamics is revealed by the evolution of the Shapiro steps with the microwave drive amplitude. Extending the Resistively Shunted Josephson junction model by including a thermal balance for the electronic bath coupled to phonons, a strong electron over-heating is obtained.
\end{abstract}

%\keywords{}
\maketitle

%\textbf{Introduction} 
A normal metal (N) coupled to two superconducting electrodes (S) constitutes a Josephson junction, that is, a device capable of sustaining a dissipationless supercurrent \cite{Likharev,Superlattices-Courtois-1999,PRB-Angers08}. The small normal state resistance and lead-to-lead capacitance of SNS junctions make these strongly overdamped in the RCSJ model \cite{Likharev}, meaning that the quantum phase dynamics is intrinsically non-hysteretic. Still, hysteresis is observed in the voltage-current ($V$-$I$) characteristics of high-critical current SNS junctions due to electronic over-heating associated to the sudden onset of dissipation when the bias exceeds the critical current \cite{PRL-Courtois-2008}. Besides, under a microwave excitation at frequency $\nu$, the junction characteristics display voltage plateaus at $V_n=n h\nu/2e$ with $n$ an integer \cite{Shapiro}. These so-called Shapiro steps are due to the phase-locking of the supercurrent oscillations at the Josephson frequency $2eV/h$ to the microwave. Shapiro steps have been frequently used for studying the phase dynamics of a variety of Josephson junctions \cite{PRL-Dubos-2001,PRL-Lehnert-1999,PRL-LeSueur-2008,PRL-Chiodi-2009,PRL-Dassonneville-2013}. This includes in particular junctions based on novel materials \cite{Science-Doh-2005,PRL-Cleuziou-2007,PRL-FuKane-2009,PRL-Jiang-2011,PRB-Dominguez-2012,CRP-Badiane-2013,PRL-Fuechsle-2009,PRB-Lombardi,NatPhys-Rokhinson,arxiv-Wiedenmann-2015,NatureNat-Pribiag} where the absence of odd steps can be the signature of topological transport. Still, the interplay between the electron over-heating and the ac Josephson dynamics has been so far overlooked.

In this paper, we discuss the response of high-critical current proximity Josephson junctions to a microwave excitation. We demonstrate the prominent role of electronic over-heating in the Shapiro steps map. Low-index steps can be masked by the switching to the resistive state. A simple model explains this behavior as well as the observation of a sharp discontinuity in the measured critical current when the ac current is increased. 

We have fabricated Nb-Au-Nb junctions using a lift-off lithography technique based on an Al-Mo metallic bilayer as a shadow mask \cite{Nanotech-Samaddar-2013}, which avoids the deterioration of Nb superconducting properties by organic contamination. Conventional e-beam lithography, followed by a dry etch of the Mo top layer and a wet etch of the Al bottom layer, produces a locally suspended Mo mask. A shadow evaporation of Au (N island) and Nb (S leads) is performed through this mask at opposite angles. The edge roughness of the structures (see Fig. 1c inset) arises from the granularity of the Mo mask. The junctions were all about 210 nm wide, the separation between the Nb electrodes ranging from 180 to 500 nm, while the Au part was about 200 nm longer in order to ensure a good overlap with each electrode. The normal-state resistance of the junctions can be related to a diffusion constant $D$ in Au of about 100 cm$^2$/s. The critical temperature of the Nb electrodes is 8.5 K. Table 1 lists the main device parameters for the different junctions investigated. 

Transport measurements were performed at temperatures down to 100 mK. Fig. \ref{DC}a-b show \emph{V-I} characteristics of a typical sample. At low temperature, a strong superconducting branch is observed (Fig. \ref{DC}a), with a large critical current $I_c$ exceeding 100 $\mu$A. This is achieved owing to the short length of the Au bridge and the high interface transparencies. Above $I_c$, the \emph{V-I} characteristic switches to the ohmic branch, characterized by the normal-state resistance $R_N$. When the current is lowered again, the $V$-$I$ characteristic remains on the ohmic branch down to the so-called retrapping current $I_r<I_c$. This pronounced hysteretic behaviour is of thermal origin and is typical of SNS junctions with a large critical current density \cite{PRL-Courtois-2008}. The retrapping current can be roughly thought of as the value of the critical current at a bias-dependent, higher electronic temperature $T^*$, determined by the thermal balance between the dissipated Joule heat and the coupling to the phonon bath. Conversely, $T^*$ is also the temperature scale above which $I_r$ and $I_c$ merge (see Fig. \ref{DC}c) and the behavior of the junction becomes non-hysteretic (Fig. \ref{DC}b).

\begin{table}[!]
%	\begin{center}
\setlength{\tabcolsep}{4pt}
\begin{tabular}{cccccccccccc}
\hline
	N$^o$ & $L$ & $R_N$ & $\, E_{\rm Th}^{\rm fit}\,$ & $\, L_{\rm eff} \,$ & $\eta$ & $I_c^0$ & $I_r^0$ & $T^*$\\
	    & (nm) & ($\Omega$) & ($\mu$eV) & (nm) & & ($\mu$A) & ($\mu$A) & (K)\\
	\hline\hline
	J1 & 225 & 2.4 & 28.1 & 474 & 0.86 & 104 & 33 & 1.9\\
	J2 & 300 & 2.1 & 23.6 & 536 & 0.89 & 78 & 45 & 1.4\\
	J3 & 180 & 1.7 & 49.6 & 390 & 0.78 & 178 & 35 & 2.6\\
	J4 & 500 & 3.6 & 9.9 & 785 & 0.67 & 14 & 2 & 1.1\\  \hline
\end{tabular}
%\end{center}
\caption{Parameters of the reported samples. In all samples, the Au strip width $W$ is 210$\pm 10$ nm, its thickness is 20 nm for sample J1 and 30 nm for samples J2 - J4, while Nb thickness is 60 nm for sample J1 and 90 nm for samples J2-J4. $L$ is the Nb-Nb distance. $R_N$ is the normal-state resistance measured at 4.2 K. $E_{\rm Th}^{\rm fit}$ and $\eta$ are the fitting parameters in Eq. 1. $L_{\rm eff}=\sqrt{\hbar D/ E_{\rm Th}^{\rm fit}}$ is the effective junction length. $I_c^0$ and $I_r^0$ the values of the critical/retrapping currents respectively at $T\to 0$ and in the absence of magnetic field and microwave excitations. $T^*$ is defined by $I_c(T^*)=I_r^0$.}
\end{table}

The temperature dependence of the critical current is displayed in Fig. \ref{DC}c. The properties of long SNS junctions depend only little on the superconducting electrodes' energy gap $\Delta$, but are mainly governed by the normal island's Thouless energy \cite{Superlattices-Courtois-1999,PRL-LeSueur-2008} $E_{\rm Th}=\hbar D / L^2 \ll \Delta$, where $D$ is the diffusion constant and $L$ the length of N. Within the Usadel equations framework and assuming $k_BT>E_{\rm Th}$, the critical current $I_c$ follows \cite{PRB-Dubos-2001}:
\begin{equation}
eR_NI_c=\eta \, E_{\rm Th}^{\rm} \frac{32}{3+2\sqrt{2}}\left[\frac{2\pi k_BT}{E_{\rm Th}^{\rm}}\right]^{3/2}\exp{\left(-\sqrt{\frac{2\pi k_BT}{E_{\rm Th}^{\rm}}}\right)}.
\end{equation}
Here the phenomenological parameter $\eta<1$ describes a reduction in the critical current related, for instance, to non-ideal interfaces. A very good fit to the data is obtained in all devices, with $\eta$ always exceeding 0.6, see Table 1. The effective length $L_{\rm eff}=\sqrt{\hbar D/ E_{\rm Th}}$ associated to the fit Thouless energy $E_{\rm Th}^{\rm fit}$ exceeds the mere separation $L$ between the electrodes. It should indeed include about twice the superconducting coherence length since Andreev reflections take place in S over such a length \cite{PRB-Dubos-2001}. The dependence of the critical current on a perpendicular magnetic field (Fig. \ref{DC}d) shows both a quasi-gaussian monotonic decay associated to dephasing and a oscillatory part arising from interference effects \cite{PRL-Bergeret-2007,PRB-Chiodi-2012}.

\begin{figure}[!t]
\includegraphics[width=\columnwidth]{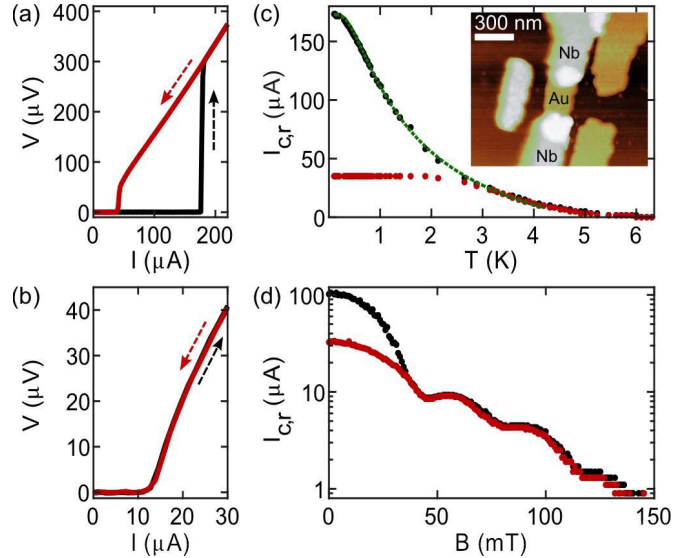}
\caption{\emph{V-I} characteristics of device J3 measured at temperatures of (a) 100 mK and (b) 4.2 K. The arrows indicate the sweeping direction of the current bias. (c) Temperature dependence of the critical current $I_c$ (black dots) and the retrapping current $I_r$ (red dots) for sample J3. The green dotted line represents a fit to Eq. (1). Inset: AFM image of a typical Nb-Au-Nb junction. (d) Magnetic-field dependence of the critical current $I_c$ (black dots) and the retrapping current $I_r$ (red dots) ($T=280$ mK, device J1). Steps in current values appearing at large magnetic field are artifacts.}
\label{DC}
\end{figure}

\begin{figure}[!t]
%\begin{widetext}
\includegraphics[width=\columnwidth]{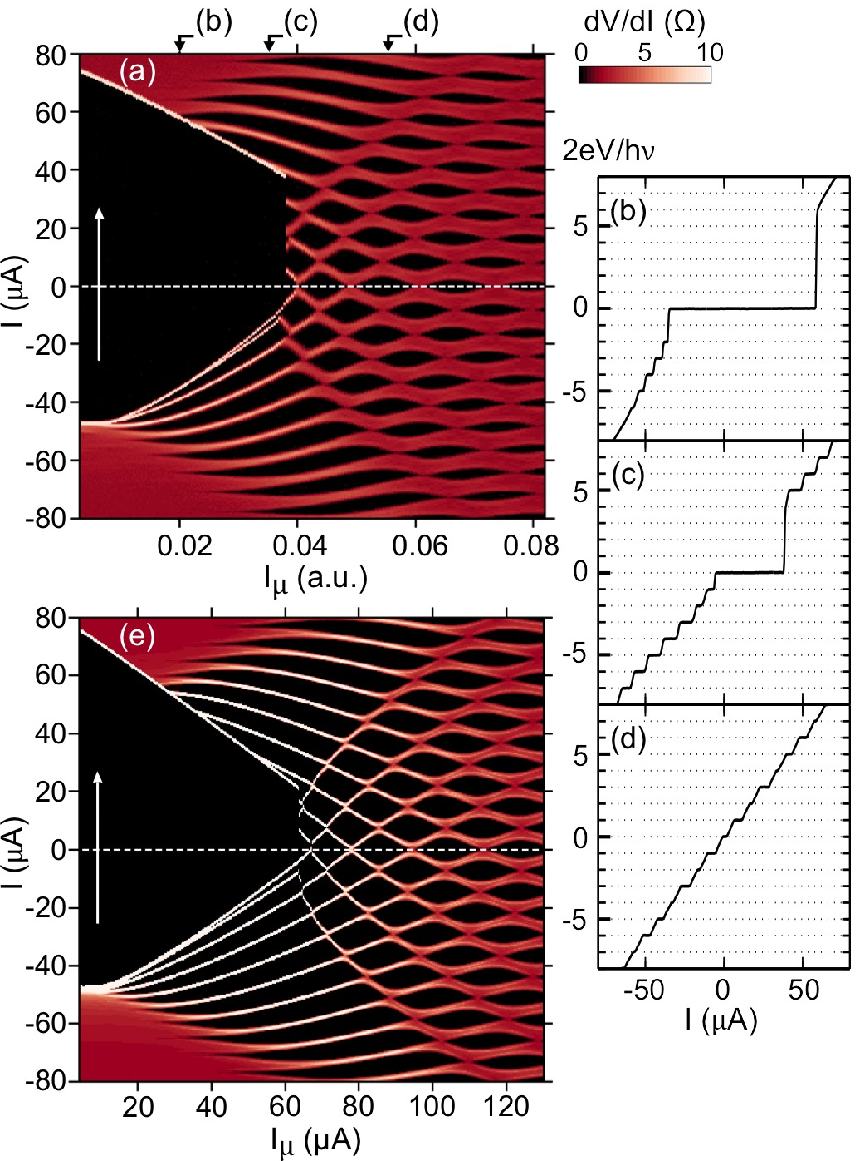}
\caption{(a) Differential resistance map as a function of the dc current bias $I$ and the microwave excitation amplitude $I_{\mu}$ for a frequency $\nu=6$ GHz (device J2, $T_{\rm bath}=100$ mK). The white arrow indicates the sweeping direction of the dc current bias. The top black arrows point to the individual \emph{V-I} curves taken at microwave drives of (b) 0.020 (c) 0.035 (d) 0.056 (a.u.). Voltage is normalized to $h\nu/2e$. (e) Calculated differential resistance map, including thermal effects.} 
\label{6GHz}
%\end{widetext}
\end{figure}

We now turn to the effect of a microwave excitation on the \emph{V-I} characteristics. Microwave signals in the frequency range 1 to 26 GHz and with a power $P_{\mu }$ were applied to the shielded cavity containing the sample. As the impedance of our samples is small compared to the estimated line impedance at high frequency, the samples are still current-biased in the microwave regime. The color map of \mbox{Fig. \ref{6GHz}a} shows the differential resistance $dV/dI$ (obtained by numerical derivation) as a function of the dc bias current $I$ and the microwave  current $I_{\mu}\propto P_{\mu}^{1/2}$ at a frequency $\nu=6$ GHz. The supercurrent branch and the Shapiro steps (up to an index exceeding 10) appear as dark regions, with zero differential resistance. At large excitation amplitudes, Fig. \ref{6GHz}a map is symmetric in $I$ and the Shapiro steps' widths oscillate with the microwave excitation amplitude, producing a well-known pattern \cite{Likharev}. At small microwave current $I_{\mu}$, the hysteresis appears through the asymmetry of the map with respect to $I$.  

Strikingly, several low-index steps do not appear in the (positive) current branch corresponding to a current increasing from zero to above the critical current $I_c$. Individual \emph{V-I} characteristics clearly demonstrate (Fig. \ref{6GHz}b-d) that the absence of these steps stems from the voltage directly jumping from zero up to about $R_NI_c$, which corresponds to the ohmic branch. Steps corresponding to voltages below $R_NI_c$ thus cannot be detected. This behavior here is clearly distinct from the discussion of recent experiments on Josephson junctions based on topological insulators, in which odd-index Shapiro steps are predicted to be absent.

Let us now consider the behavior of the critical current as a function of the microwave current amplitude $I_{\mu}$. In a current bias picture, the microwave irradiation adds adiabatically an oscillatory excursion $I_{\mu}$ to the bias current $I$, so that the current oscillates between $I - I_{\mu}$ and $I + I_{\mu}$ \cite{PRL-Chiodi-2009}. In a quasi-static regime, switching to the resistive state then occurs at a lower critical current $I_c-I_{\mu}$. Once the junction has switched, it remains in the resistive state due to hysteresis. One expects therefore a linear suppression of $I_c$ with increasing $I_{\mu}$, as seen in Fig. \ref{6GHz}a. We attribute the slight downward deviation from linear behavior to a small increase in the electronic temperature at high microwave power. On the retrapping branch, a similar decay of the retrapping current $I_r$ with the microwave current $I_{\mu}$ is observed. 

In order to provide a quantitative description, we need to consider the energy relaxation from the normal metal electronic population. In the present temperature range, electron-phonon scattering is the dominant mechanism. The related coupling power between electrons at a temperature $T_e$ and phonons at a temperature $T_{ph}$ is $P_{e-ph}=\Sigma U (T_e^5-T_{ph}^5)$, where $\Sigma$ is the material-dependent coupling constant and $U$ is the metal volume. Considering a retrapping temperature $T^*$ of 1 to 3 K, the related rate $\tau_{\rm e-ph}^{-1}\approx 0.16\times T^3$ GHz in Au \cite{PRB-Echternach-1992} is in the low GHz range ($\approx 0.4$ GHz in J2 at $T^*$ = 1.4 K for instance). In most of the frequency range investigated here, the thermal relaxation is thus slow compared to the microwave ($\tau_{\rm e-ph}^{-1} < \nu$) so that the electronic temperature can be considered as almost constant with time at a given ($I$,$I_{\mu}$) bias point. 

We consider a Resistively Shunted Junction (RSJ) model \cite{Likharev} with a current bias. The time-dependent current $i(t)=I+I_{\mu} \sin (2 \pi \nu t)$ through the junction is considered as the sum of the ohmic current $v/R$ and the Josephson current $I_c \sin \varphi$, where $\varphi$ is the phase difference across the junction:
\begin{equation}
i(t)= I +I_{\mu}\sin(2 \pi \nu t)= I_c \sin \varphi + v/R.
\label{current}
\end{equation}
The time-dependent voltage $v$ relates to the time-derivative of the phase as $v(t)=\hbar {\dot \varphi}/2e$ from the second Josephson relation. From Eq. (\ref{current}), the phase dynamics can be modeled as that of a massless particle of position $\varphi$ in a {\it tilted washboard} potential $U(\varphi)=-\hbar [I_c \cos \varphi+i(t)]/2e$. The potential slope is proportional to the current bias $i(t)$: its average is thus determined by $I$ and it oscillates with an amplitude given by $I_{\mu}$. For large enough $I$ or $I_{\mu}$, the particle rolls down the slope. The Shapiro steps at voltage values $V_n=n\, h \nu/2e$ correspond to the particle hopping down by $n$ minima during one microwave period.

We can write the instantaneous Joule power:
\begin{equation}
p(t) = i(t) \cdot v(t) = I_c  \frac{\hbar}{2e}{\dot \varphi} \, \sin\varphi + \frac{1}{R}\left(\frac{\hbar}{2e}{\dot \varphi}\right)^2.
\label{power}
\end{equation}
The first term relates to the change in the Josephson energy. It is zero in average and does not contribute to the average power $P$ dissipated over one cycle. Only the second term, which is the Joule power across the junction resistance, dissipates. It can be non-zero in average even though the average voltage $V$ is zero.

At this point, we now introduce a heat balance where the dissipated power $P$ is balanced by the electron-phonon coupling power $P_{e-ph}$. The related temperature elevation acts on the phase dynamics through the temperature dependence of the critical current following Eq. (1). Using Eqs. (\ref{current}) and (\ref{power}) and taking the volume $U$ as a free parameter, we have numerically calculated the time-dependence of the phase, the dissipated power and the ensuing electronic temperature $T_e$ for every ($I$,$I_{\mu}$) bias point, which gives access to the related dc voltage drop $V$. Fig. \ref{6GHz}e shows the calculated differential resistance for device J2's parameters. For the best fit, the volume $U$ was chosen as 10 times the physical volume. This can be explained by both the inverse proximity effect in the leads in the vicinity of the N-S interface and the thermal conductance of the leads between the N island and the N metal regions of the leads. A semi-quantitative agreement for the differential resistance between Fig. \ref{6GHz}a and \ref{6GHz}e is readily seen.

\begin{figure}[!t]
%\begin{widetext}
\includegraphics[width=\columnwidth]{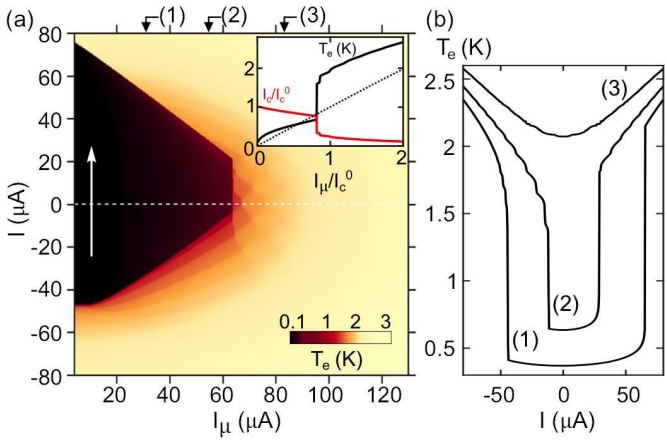}
\caption{(a) Calculated electronic temperature $T_e$ map corresponding to the situation of \mbox{Fig. 2}, \mbox{i.e.} for device J2 at $T_{\rm bath}=100$ mK. Inset: Line cut of this map showing the electronic temperature at zero dc bias current as a function of the microwave current and calculated value of the critical current at this electronic temperature. The dashed line indicates the $I_c(T_e) = I_{\mu}$ correspondence. (b) Line cuts of the map (a) showing the dependence of the electronic temperature as a function of the dc bias current $I$ at different values of the microwave current $I_{\mu}$ indicated by arrows at the top of the map.} 
\label{Temp}
%\end{widetext}
\end{figure}

The associated temperature map of Fig. \ref{Temp}a highlights the importance of dissipation in the ac phase dynamics in SNS junctions. Strikingly, the electronic temperature varies significantly as a function of the microwave current bias: the temperature increases from the bath temperature of 0.1 K up to above 2 K. On the first Shapiro step, the temperature is already of about 1 K. Even for zero dc current $I$ and hence zero average voltage $V$, electrons in N are significantly overheated at large microwave drives, see Fig. \ref{Temp}a inset. The Shapiro steps structure appears also on the temperature map, as can be seen in Fig. \ref{Temp}a and more clearly in Fig. \ref{Temp}b as wiggles in every curve, especially the curve (2). 

Both the data (Fig. \ref{6GHz}a) and the calculations (Fig. \ref{6GHz}e) exhibit a sudden drop of the critical current $I_c$ as the retrapping current $I_r$ approaches zero at a microwave current $I_{\mu}=I^*$ (of about 0.04 in Fig. 2a). As shown in Fig. \ref{Temp}a inset, the microwave amplitude $I_{\mu}$ at this point is approximately equal to the critical current $I_c(T_e)$ at the electronic temperature at zero bias, so that no supercurrent can be established due to the electronic overheating at this point of the V-I characteristics. The numerical solution of the phase dynamics illustrates this precisely: while at an ac drive below $I^*$ the particle oscillates in a single washboard valley, above $I^*$ it hops back and forth between two valleys \cite{SuppMat}. Let us note that this feature would not appear if the microwave amplitude would be swept at a fixed (and low) dc bias current. 

We have obtained a good agreement between experiment and calculation at every investigated microwave frequency up to 26 GHz. At higher frequency, as dissipation increases with excitation frequency $\nu$ as $\nu^2$, the electron over-heating is much increased in the superconducting state ($V=0$). The two currents $I_c(I_{\mu})$ and $I_r(I_{\mu})$ are thus seen to merge before $I_r$ approaches zero \cite{SuppMat} and no discontinuity is observed.

\begin{figure}[!t]
\includegraphics[width=0.99\columnwidth]{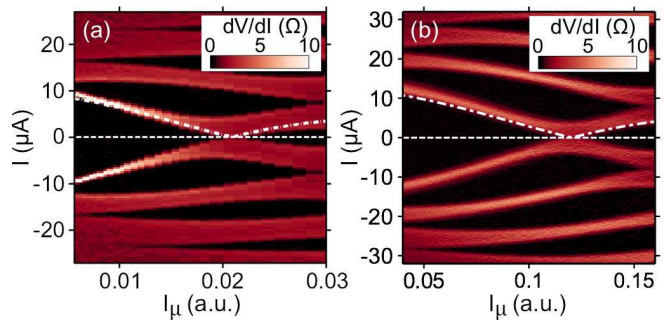}
\caption{Differential resistance maps as a function of {\it I} and $I_{\mu}$ for $\nu=6$ GHz in non-hysteretic conditions, (a) at base temperature and under a magnetic field of 100 mT (device J2), and (b) at a bath temperature of about 4.2 K (device J3, no magnetic field). The white lines represent the usual Bessel-function-like amplitude of the step $n=0$.}
\label{nonhysteretic} 
\end{figure}

The interplay of electron over-heating and ac Josephson dynamics discussed here is prominent in junctions that are hysteretic in dc. Indeed, when driving our same devices to non-hysteretic conditions, either by applying a magnetic field ({\mbox Fig. \ref{nonhysteretic}a}) or increasing the temperature ({\mbox Fig. \ref{nonhysteretic}b}), the usual Shapiro pattern is recovered \cite{Likharev}. This is evidenced by the good agreement of the width of the $n=0$ step with the standard Bessel function expression (white dotted lines in {\mbox Fig. \ref{nonhysteretic}a} and b). The crossover between the overheating regime and the isothermal behaviors depends not only on the junction critical current amplitude but also on the coupling to the thermal bath.

In summary, we have evidenced that electron over-heating is of paramount influence in the microwave response of Josephson junctions. Exploring the microwave response of Josephson junctions involves variable electronic temperatures, which is of uttermost importance for the complete analysis of devices based on new materials like topological conductors.

\begin{acknowledgments}
We acknowledge financial support from the ANR contract "Nanoquartets" and the LANEF project "UHV-NEQ". Samples were fabricated at the Nanofab platform at CNRS, Grenoble. We thank A. Nabet and D. van Zanten for help in the experiments, S. Samaddar and J. P. Pekola for discussions.
\end{acknowledgments}

\clearpage
\widetext
\begin{center}
\textbf{\large Supplemental Materials: Interplay Between Electron Over-Heating and ac Josephson Effect}
\end{center}
%%%%%%%%%% Merge with supplemental materials %%%%%%%%%%
%%%%%%%%%% Prefix a "S" to all equations, figures, tables and reset the counter %%%%%%%%%%
\setcounter{equation}{0}
\setcounter{figure}{0}
\setcounter{table}{0}
\setcounter{page}{1}
\makeatletter
\renewcommand{\theequation}{S\arabic{equation}}
\renewcommand{\thefigure}{S\arabic{figure}}
\renewcommand{\bibnumfmt}[1]{[S#1]}
\renewcommand{\citenumfont}[1]{S#1}
%%%%%%%%%% Prefix a "S" to all equations, figures, tables and reset the counter %%%%%%%%%%
\newcommand{\ep}{\epsilon}
\newcommand{\vep}{\varepsilon}
\newcommand{\hc}{{\rm \;h.\,c.\;}}
\newcommand{\sign}{\mathop{\mathrm{sign}}\nolimits}
\renewcommand{\Im}{\mathop{\mathrm{Im}}\nolimits}
\renewcommand{\Re}{\mathop{\mathrm{Re}}\nolimits}
 %\usepackage[style=nature]{biblatex}

%\documentclass[aps,prl,twocolumn,groupedaddress,showpacs]{revtex4-1}

% You should use BibTeX and apsrev.bst for references
% Choosing a journal automatically selects the correct APS BibTeX style file (bst file), so only uncomment the line below if necessary.
%\bibliographystyle{apsrev}
%\usepackage{graphicx}
%\usepackage{float}
%\begin{document}
% Use the \preprint command to place your local institutional report number in the upper righthand corner of the title page in preprint mode.
% Use the 'preprintnumbers' class option to override journal defaults  to display numbers if necessary
%\preprint{}

%\title{Electron Over-Heating Masks Shapiro Steps in Proximity Josephson Junctions}
\title{Interplay of Electron Over-Heating with ac Josephson Effect}

% repeat the \author .. \affiliation  etc. as needed  \email, \thanks, \homepage, \altaffiliation all apply to the current author. Explanatory text should go in the []'s, actual e-mail address or url should go in the {}'s for \email and \homepage.
% Please use the appropriate macro foreach each type of information \affiliation command applies to all authors since the last
% \affiliation command. The \affiliation command should follow the other information
\author{A. De Cecco$^{1,2}$, K. Le Calvez$^{1,2}$, B. Sac\'ep\'e$^{1,2}$, C. B. Winkelmann$^{1,2}$, H. Courtois$^{1,2}$}
%\thanks{}
\affiliation{$^{1}$Universit\'e Grenoble Alpes, Institut N\' eel, 25 Avenue des Martyrs, 38042 Grenoble, France}
\affiliation{$^{2}$CNRS, Institut N\' eel, 25 Avenue des Martyrs, 38042 Grenoble, France}

%Collaboration name if desired (requires use of superscriptaddress option in \documentclass). \noaffiliation is required (may also be used with the \author command).
%\collaboration can be followed by \email, \homepage, \thanks as well. \collaboration{}

\date{\today}
\maketitle
In this supplemental materials part, we provide some more details about the numerical calculation of the differential resistance map under microwave irradiation, as well as the calculation results supporting the explanation for the discontinuity observed in the differential resistance map. We also discuss additional experimental data obtained at different values of the microwave frequency.

%\pacs{74.50.+r, 74.45.+c}
%\keywords{}

\section{Numerical calculations}
 
Our model is an extension of the RSJ model with the temperature dependence of the critical current (Eq. (1) of the main text), the electronic temperature $T_e$ in N being governed by the thermal balance between the Joule power and the coupling to phonons. We here neglect other heat transport channels out of N, which will eventually lead to a somewhat lower temperature. 

\begin{figure}[b]
\includegraphics[width=0.5\columnwidth]{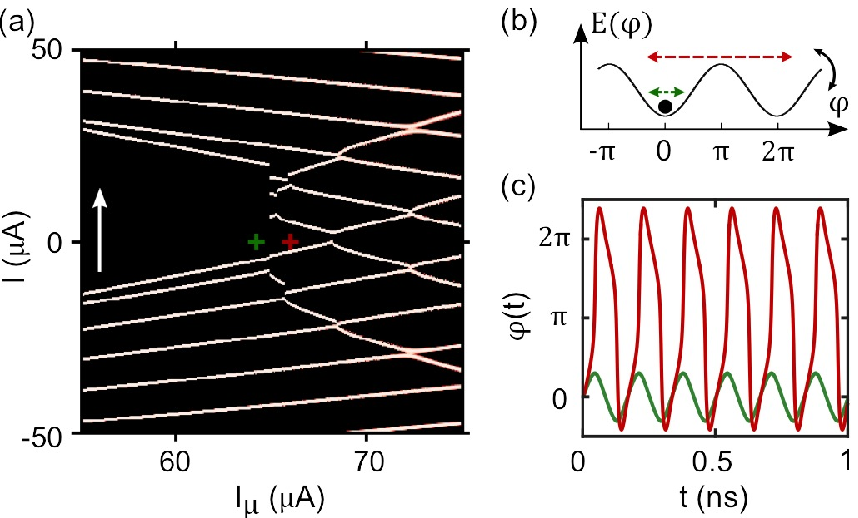}
\caption{(a) Zoom of the differential resistance map as a function of $I$ and $I_{\mu}$ at $\nu=6$ GHz (device J2, $T_{bath} = 100$ mK) shown in Fig. 2e in the main paper. (b) Schematics showing the extent of the phase excursion at the two bias points. (c) Calculated time dependence of the phase at the bias points indicated by the crosses in (a). The dc current is zero in both cases.}
\label{Zoom} 
\end{figure}

A sharp discontinuity in the differential conductance map at 6 GHz is described in the main text, see Fig. 2 of the main paper. Fig. \ref{Zoom} shows both a zoom of the map and the time-dependence of the phase at two points at the same dc current bias, but at an ac current signal amplitude just below and above the threshold. While the mean voltage is zero in every case, the phase's excursion is larger in amplitude at larger ac signal. It exceeds 2$\pi$ in the latter case, meaning that the effective particle travels over two neighboring minima of the potential landscape. The related dissipation makes the electronic temperature rise and the effective critical current drop, resulting in a sharp change in the differential conductance map.

\begin{figure}[!t]
\includegraphics[width=0.5\columnwidth]{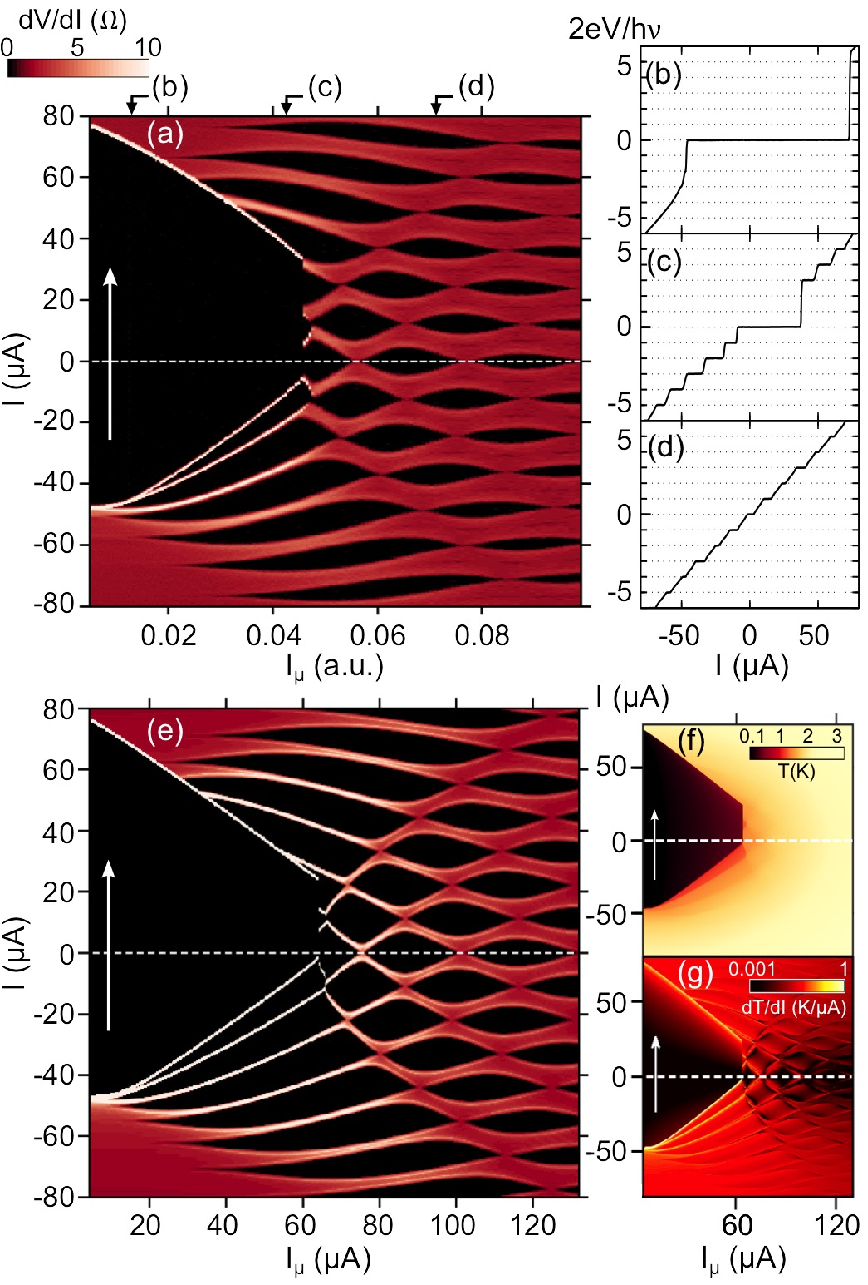}
\caption{(a) Differential resistance map as a function of the dc current bias $I$ and the microwave excitation amplitude $I_{\mu}$ for a frequency $\nu=8.8$ GHz (device J2, $T_{\rm bath}=100$ mK). The white arrow indicates the sweeping direction of the dc current bias. The top black arrows point to the individual \emph{V-I} curves taken at microwave drives of (b) 0.010 (c) 0.043 (d) 0.072 (a.u.). Voltage is normalized to $h\nu/2e$. (e) Calculated differential resistance map, including thermal effects, see text for details. The related calculated temperature is displayed in (f), its derivative with respect to bias current is in (g).} 
\label{8GHz} 
\end{figure}

\begin{figure}[!t]
\includegraphics[width=0.5\columnwidth]{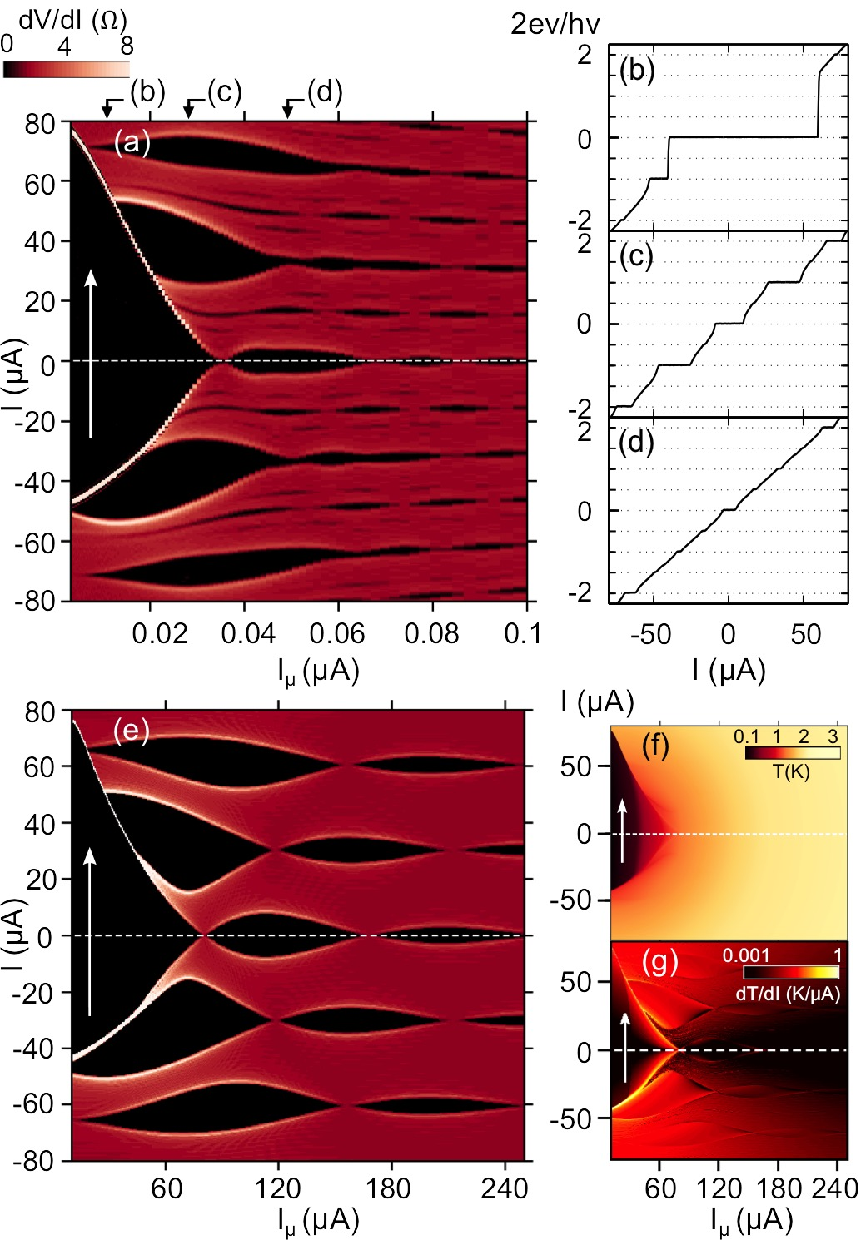}
\caption{(a) Differential resistance map as a function of $I$ and  $I_{\mu}$ at $\nu=24.2$ GHz (device J2, $T= 100$ mK). The white arrow indicates the sweeping direction of the dc current bias. The top black arrows point to the individual \emph{V-I} curves taken at microwave drives of (b) 0.010 (c) 0.028 (d) 0.049 (a.u.). Voltage is normalized to $h\nu/2e$. (e) Calculated differential resistance map, including thermal effects, see text for details. The related calculated temperature is displayed in (f), its derivative with respect to bias current is in (g).} 
\label{24GHz} 
\end{figure}

The agreement with experimental data at the same frequency is remarkable at every microwave frequency investigated. Fig. 2 in the main paper displays data and calculation at 6 GHz, while Fig. \ref{8GHz} and Fig. \ref{24GHz} in this supplementary materials display similar information at 8.8 GHz and 24.2 GHz.

\begin{figure}[!b]
\includegraphics[width=0.5\columnwidth]{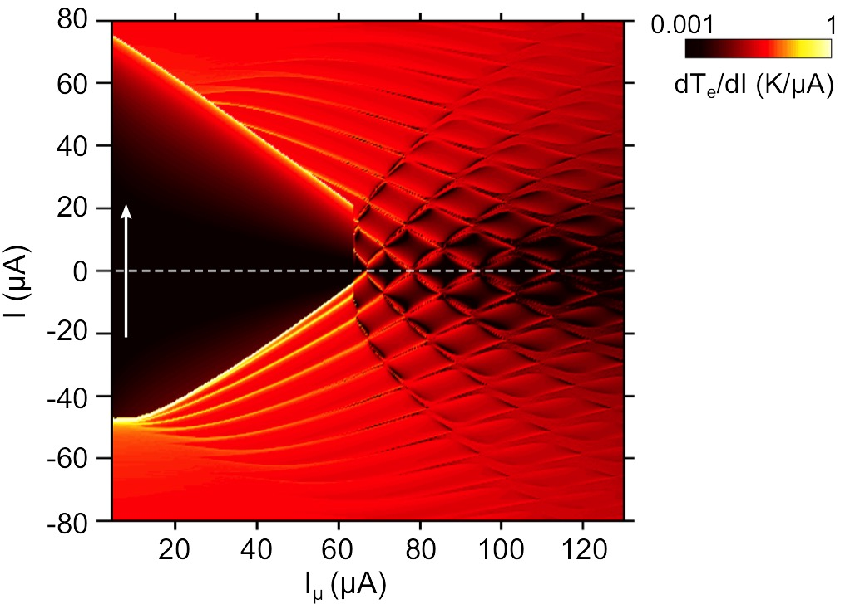}
\caption{Calculated map of the temperature derivative with respect to the current bias, including thermal effects, see text for details, corresponding to the case of Fig. 2 of the main paper, \mbox{i.e.} for sample J2 at a bath temperature of 100 mK.} 
\label{Derivative} 
\end{figure}

Here we also provide the map of the temperature derivative with respect to the current bias, see Fig. \ref{Derivative}. This plot highlights the appearance in the temperature map of the Shapiro steps structure.

\section{Behavior at higher frequency}

In the higher frequency regime, we observe the appearance of fractional Shapiro steps at voltages $V_{n,m}=(n/m)\,h\nu/2e$, where $n$ and $m$ are integers. This appears as thin zero-differential resistance regions at intermediate positions compared to the integer steps in Fig. \ref{24GHz}a, and short steps in Fig. \ref{24GHz}b-d. Fractional Shapiro steps were already observed in similar devices \cite{PRL-Lehnert-1999b,PRL-Dubos-2001b,PRL-Fuechsle-2009b}. Multiple Andreev Reflections as well as non-thermal out-of-equilibrium energy distribution function can contribute to this phenomena. As they are not included in our model, fractional steps are absent from the calculation results.

At higher frequency, the dissipation at a given amplitude $I_{\mu}$ is larger. The dominating effect of the microwave irradiation on the electronic population appears in Fig. \ref{24GHz} through the merging of the critical current and the retrapping current above an intermediate level of ac current bias $I_{\mu}>0.02$. As a consequence, no discontinuity is observed in the differential resistance map. Still, the n = 1 step is missing at low microwave excitation.

 \end{document}